\documentclass[letterpaper,twocolumn,10pt]{article}
\usepackage{usenix2022_SOUPS}

\usepackage{tikz}
\usepackage{float}
\usepackage{amsmath}
\usepackage{graphicx}
\usepackage{acronym}
\usepackage{balance}
\usepackage{placeins}

\acrodef{CSIRT}{Cyber Security Incident Response Team}
\acrodef{CISO}{Chief Information Security Officer}
\acrodef{EA}{Enterprise Architecture}
\acrodef{D-D}{Decisions \& Disruptions}
\acrodef{D-D2}[D-D 2]{Decisions \& Disruptions 2: Decide Harder}

\begin{document}

\date{}

\title{\Large \bf Decisions \& Disruptions 2: Decide Harder\\A custom cyber security incident response exercise}

\def\plainauthor{Shreeve et al.}

\author{
  {\rm Benjamin Shreeve}\\
  University of Bristol
  \and
  {\rm Joseph Gardiner}\\
  University of Bristol
  \and
  {\rm Joseph Hallett}\\
  University of Bristol
  \and
  {\rm David Humphries}\\
  City of London Police
  \and
  {\rm Awais Rashid}\\
  University of Bristol
} 

\maketitle
\thecopyright

\begin{abstract}
Cyber incident response is critical to business continuity---we describe a new exercise that challenges professionals to play the role of \ac{CISO} for a major financial organisation. Teams must decide how organisational team and budget resources should be deployed across \ac{EA} upgrades and cyber incidents. Every choice made has an impact---some prevent whilst others may trigger new or continue current attacks. We explain how the underlying platform supports these interactions through a reactionary event mechanism that introduces events based on the current attack surface of the organisation. We explore how our platform manages to introduce randomness on top of triggered events to ensure that the exercise is not deterministic and better matches incidents in the \emph{real world}. We conclude by describing next steps for the exercise and how we plan to use it in the future to better understand risk decision making.
\end{abstract}

\section{Introduction}

Major cyber security incidents regularly disrupt businesses, and in extreme circumstance have even bankrupted them. We have created a major new incident response exercise to help businesses. We have worked with specialists from law enforcement and major financial organisations to create an exercise that challenges teams to handle a major, responsive cyber security incident. The aim of the exercise is to expose senior managers and incident response teams to the time, resource and political pressures they will encounter whilst handling a major crisis whilst at the same time gathering granular decision-making data to inform cyber response handling research. 

This work builds upon a freely available previous game released under a CC-BY-NC license: \acf{D-D}~\footnote{\url{decisions-disruptions.org}}. \ac{D-D} is a highly successful tabletop exercise utilised by police forces across the UK and businesses across the world. It was designed to explore how people make risk decisions around cyber-physical infrastructures. Whilst D-D has provided valuable practical and research insights \cite{Frey2017,Shreeve2020soifmrblue,Shreeve2020Thebestlaidplans,shreeve-2022-how-managers}, we argue that it is limited by a deterministic mechanism, fixed and tied to one sector. We, therefore, propose a new game: \acf{D-D2} which provides an extensible engine for risk decision making exercises that incorporates randomness and more complicated threat relationships that can be targeted towards any industry, rather than just for critical national infrastructure. This paper introduces our proposed new game \ac{D-D2} and the new features it incorporates.

\section{Related Work}
Cyber security exercises are a popular way of raising awareness of the subject matter. A number of exercises have been developed specifically as part of University courses (e.g,~\cite{bock-2018-kingofthehill, morelock-2018-authenticity}). However, such exercises tend to have a relatively narrow scope related to the content of specific University courses and they are rarely validated with or used outside of academia. 

Other existing exercises have been created to help raise awareness of cyber security issues in industry (e.g.,~\cite{denning-2013-control-alt-hack,gondree-2013-valuing,shostack2014elevation,Frey2017,beckers2016serious,Hart-Riskio-a-serious}). All of these exercises are tabletop exercises and with the exception of Frey et al.~\cite{Frey2017} all use card-based mechanisms. Whilst the card-based mechanisms are valuable for raising awareness they often limit how well exercises can reflect real-world scenarios. For example, such mechanisms rely on a mixed deck of cards (or several decks) which are then drawn from at random to provide game events. As such it is hard to such mechanisms to capture the way that events in the real-world may be related with one event causing another to occur. This is not to say they are not of value, but they often emphasise learning of specific aspects rather than emulation of scenarios. For example, Hart et al.~\cite{Hart-Riskio-a-serious} introduce the Riskio serious game---their exercise is aimed at non-technical participants and challenges them to consider potential threat vectors and then identify possible countermeasures. This provides a no doubt valuable learning experience, but is not the same as exposing participants to an emulation of decision-making under pressure which we is the aim of our exercise. Hart et al.~\cite{hart2021motens} have taken lessons learned from the development of Riskio~\cite{Hart-Riskio-a-serious} and used it, along with a careful analysis of a wide range of cyber security games (including \ac{D-D}), to create the MOTENS design model for serious games. This model suggests that the most effective cyber security games include: \textbf{M}ultiple modes of learning---exposing players to a wide range of cyber security aspects; \textbf{O}wnership Self-Learning---Providing a range of options to help meeting learning objectives;\textbf{T}heory---that supports the design;\textbf{E}nvironment---creating an appropriate environment where people feel they can learn;\textbf{N}egotiation---moving toward a coaching and problem-based learning style;\textbf{S}elf learning---enabling participants to build upon their base knowledge.

Frey et al.~\cite{Frey2017} provide a different approach---their exercise \emph{Decisions \& Disruptions} provides teams with a Lego representation of a hydro-electric company with an plant (operations) site and separate office site. Teams play through 4 rounds, investing a budget each round to implement security controls on these sites. At the end of each round they discover what cyber attacks have befallen the business as a result of their choices. The game mechanism makes it clear to players that there is a direct correlation between the investment choices they have made and the events they have suffered. This is a significant improvement in terms of realism, enabling more complex attack scenarios to be developed. However, there are limitations to the mechanism created by Frey et al.~\cite{Frey2017}, the game mechanism has hard-coded paths through it (e.g.,~event A will occur if security control B has not been purchased by round 3). This means that teams cannot replay the exercise as they will always experience the same set consequences for their actions. 

We seek to build on the success of \ac{D-D} by developing an exercise where the players affect not only the landscape in which they play but also the consequences of their choices. An exercise where events that have occurred and choices that have been made increase the likelihood of (or even directly cause) specific events to occur, and where events in turn can change the landscape. We work closely with \acp{CISO} and \acp{CSIRT} to develop a unique, replayable exercise, inspired by \ac{D-D} that challenges teams to make cyber security decisions under the time, resource and political pressures of a series of unfolding cyber incidents.

\section{The Exercise}
The exercise challenges teams to help the \ac{CISO} of a fictitious financial organisation handle a \emph{very bad} cyber security week---each day of the week is represented by a 20 minute time-limited round. Each day, a series of cyber-related events will occur, which can be handled in various ways. Players have to decide which tasks the \ac{CISO}'s team will tackle each day to stay on top of what's happening to their company.

Teams have the ability to affect the overall attack-surface of the business and therein the type and effectiveness of possible attack vectors used against the business by updating assets through actions such as patching and staff training. How attacks (or symptoms) are handled affect whether attack(s) are prevented, continue or whether new related attacks are triggered. The exercise is designed to be played by \ac{CSIRT}, \acp{CISO} and senior executives. 

\subsection{A Reactionary Event Mechanism}
\label{sec:reactionary event mechanism}
\begin{figure}[t]
    \centering
    \includegraphics[width=0.80\columnwidth]{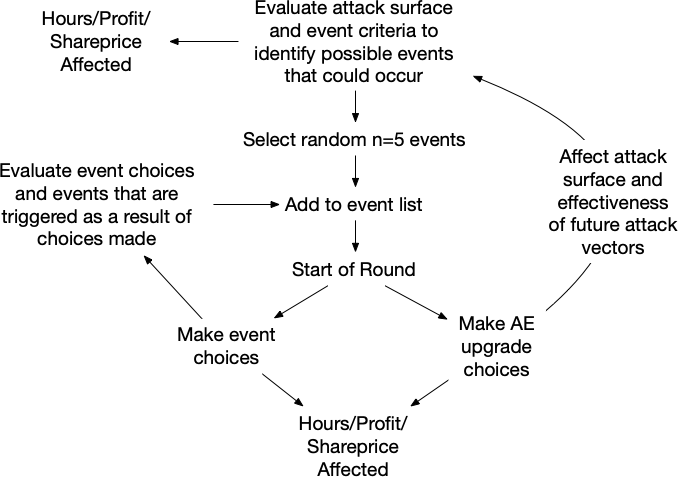}
    \caption{Summary of the different mechanisms in the game and the outcomes that can happen.}
    \label{fig:mechanism_summary}
\end{figure}

\begin{figure}[t]
    \centering
    \includegraphics[width=.73\columnwidth]{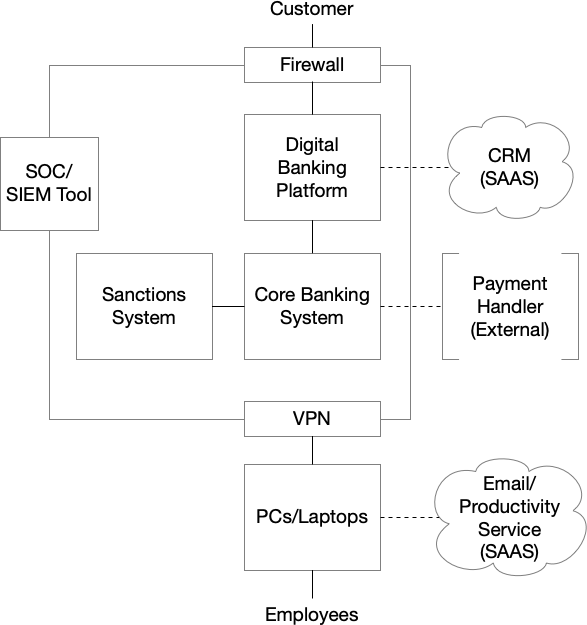}
    \caption{Financial Organisation enterprise architecture}
    \label{fig:EA}
\end{figure}

In order to be able to represent the complexities of real-world decision-making a complex and reactionary event mechanism was developed (see figure~\ref{fig:mechanism_summary}). Teams are tasked with protecting the \ac{EA} of the business (see figure~\ref{fig:EA}). They are able to affect the company's attack surface by investing staff hours each day into a wide range of possible upgrades (and in some cases by sacrificing profit). The attack surface is evaluated at the start of each round and used to identify which events can occur (out of a library of 120 possible events). Of those that can occur, 5 are randomly selected and added to the event list for the next round. Events can be specified to only occur given a particular set of criteria---such as a particular EA upgrade having been purchased or a previous event having occurred. Each choice that a team makes has an associated cost in terms of staff hours and company profitability. Events have been carefully designed in tandem with cyber security law enforcement officers to provide a realistic representation of the threats facing financial organisations. The range of events reflect the major MITRE ATT\&CK~\footnote{\url{https://attack.mitre.org/matrices/enterprise/}} areas highlighted through interviews with Chief Information Security Officers (CISOs) in major financial organisations as problematic. 

For many choices there are also consequences. These are either explicit feedback to teams as to the impact of their choices---including additional impact to hours/profitability/shareprice---or an in-game consequence which can include triggering other events in the future. These performance indicators were flagged up through interviews with CISOs and board members of major financial organisations as vital indications of performance during a cyber incident~\cite{TOSUN2021101795,Agrafiotis-a-taxonomy}. These triggered events are then added to the event list for the next round (possibly bypassing any criteria evaluation). Events can also trigger other events to occur if they are ignored; for example if a team decides an event is not important in one round it can still have an impact on their next round by queuing up an event to penalise their neglect. Some events are designated \emph{`on-draw'} events---they affect the round as soon as they are drawn, deliberately introducing variance to the game. For example, an event may tell a team that a member of the \ac{CSIRT} is ill that day; they therefore will start the round 8 hours down, with only 72 hours available to them to utilise. These \emph{`on-draw'} events can force hours, profitability, shareprice and even make \ac{EA} upgrade choices for teams before a round starts. 

For example, as part of the game an event may occur where a staff member reports finding a USB stick on the ground.  The CISO can decide whether to ignore it, to forensically analyse it (at a cost of time) or to destroy the device.  Each decision will have impacts (and could in turn lead to more events occurring.  To visualise these kinds of decisions we created an automated tool to create attack trees from the engine's database of events and explore what happens. This gives a quick pictorial guide to the consequences of any decision in the game and the threat landscape of any particular configuration.


\subsection{Resources}

Teams have to negotiate various resource constraints. Firstly, they have to contend with the challenge that a day has a finite length---in this case a configurable 20 minute limit. If they fail to utilise their resources effectively in that time then the game automatically moves forward to the next day. The primary unit of daily resources are the number of \emph{`hours'} available to the team each day---that is, how many of staff in the \ac{CSIRT} team that they are managing are available that day (the exercise assumes that there are 10 staff members, each with a maximum 8 hours per day, 80 hours in total per day). These hours are then allocated to event/\ac{EA} actions. 

The businesses performance is represented by two broad financial metrics that are both affected by choices made and consequences. Firstly, the team have to manage the overall long-term \emph{`projected profit'} for the business---this starts at \pounds{}500,000. However, it can be reduced through loss of business or fines that occur as a result of choices made. Teams can also choose to reinvest some of this projected profit into the business through certain \ac{EA} upgrades or event choices which have associated financial costs. Secondly, teams have to consider the short-term performance of the business in the form of its \emph{`share price'}---which can be affected both positively and negatively as a result of choices made from an initial value of \pounds{}100. If the players actions result in the \emph{`projected profit'} or \emph{`share price'} reaching zero then the game ends. 

\subsection{Exercise Management}
The exercise comprises two parts: a game master tool that tracks and evaluates the state of play and an associated physical card deck (see figure~\ref{fig:physical-cards}). The game master utilises a digital tool, written in Java, that reads a database containing details on the possible events (and their relationship to one another) as well as the assets and upgrades that possible for the \ac{EA}. Each event/choices and asset/upgrade are also printed to a deck of cards providing session flexibility. 

\subsection{Typical Play Through}
At the start of each \emph{day} (round) the exercise interface updates to reflect the status of the business (see example in appendix~\ref{appx:exerciseUI}). 

This right hand panel consists of a list of possible actions that can be undertaken to help improve the cyber security of their assets. The left hand panel shows a list of the most pertinent cyber events that have hit the business in the last 24 hours and that need resolving (a minimum of 5 occur each day) plus any that have yet to be actioned in some way that are remaining from previous rounds.

The \ac{EA} upgrades (see example in appendix~\ref{appx:exerciseUI}) can be purchased at any point during a round and affect the attack surface of the game in the next round. 

Each event presents presents teams with several choices about how they might choose to resolve each of these events (see example in appendix~\ref{appx:exerciseUI}). All events by default include `ignore this round' which defers making a choice for that event to the next round. All choices are final---once a choice has been made (other than ignore), then it cannot be undone.

Teams work together to identify which events and \ac{EA} upgrades they should dedicate resources towards in the round. As they make each choice the game provides them with instant feedback which may result in their hours, profitability or share price being immediately hit, which may in turn then limit the choices they had planned. When the round concludes the game utilises the reactionary event mechanism logic (see section~\ref{sec:reactionary event mechanism} and figure~\ref{fig:mechanism_summary}) to evaluate which events will occur in the next round (as a result of changes the team have made to the businesses attack surface), or that the team have directly triggered (because of choices made in during the round). 

The exercise continues this way through multiple rounds with teams starting each round with at least 5 events to address and a starting number of hours for the day affected by `on-draw' events. The exercise concludes either when the round limit is reached (7 rounds) or when one of the failure criteria are triggered. The game has 3 failure criteria: the profitability of the business can reach zero, the share price can reach zero or the team can have 10 or more concurrent events open. We consider a team that has failed to resolve (or has triggered) 10 (or more) events to have reached a point of event saturation that could never be resolved in the real world. The extent to which consequences of event choices affect the profitability and shareprice of the organisation vary based on the magnitude of the event. Some consequences may have a positive impact whilst others may be severe, or even sufficient to bankrupt the organisation. 

\section{Design Choices}

We have created a mechanism that can identify if an attack is viable based upon not only the current state of the organisations attack surface, but also on whether specific previous events have occurred and particular choices made. This means that events that are presented are far more representative of real-world scenarios.

We have taken this further by treating certain defensive approaches as \emph{`resistances'}. Defensive mechanisms, like \emph{firewalls} and \emph{antivirus}, can never be 100\% effective. Instead, our game mechanism captures the current success rate of these mechanisms (for example the Firewall is only 60\% effective) and then uses these values to establish if an attack that could be prevented by a firewall will occur by incorporating chance. We also provide teams with the ability to improve these resistances by investing in specific \ac{EA} upgrades---and punish \emph{negligence} for not doing so in a timely manner. In doing so we create an exercise where there is no longer a 1:1 relationship between events and defences and instead there are combinations of defences that can limit the likelihood of specific attack vectors being exploited---just like in the real world. 
The tool is itself also extensible. The events for a given game are taken from a database and can be rewritten for different sectors or markets. Penalties for different events and how events interlink is configurable allowing for a wide range of variations of the game to be created with relative ease.

Our design choices fulfil Hart et al.'s~\cite{hart2021motens} MOTENS pedagogical design framework for serious cyber games: \textbf{M}ultiple Modes of Learning: The mechanism enables participants to experience not only a range of events, but also to explore the relationship between \ac{EA} and system security and attacks. \textbf{O}wnership and Self-Learning: The underlying platform is flexible, allowing sessions to be tailored to specific learning objectives. \textbf{T}heory: The overarching game principle is informed by experiential learning theory~\cite{kolb2014experiential}. \textbf{E}nvironment: The exercise environment is designed so teams have more forgiving opening rounds to help them familiarise themselves with the exercise mechanics and expectations. \textbf{N}egotiation: The promotion of shared decision-making amongst participants and the immediacy of feedback through the reactionary event mechanism move learning from static presentation to an immersive exploratory experience. Finally, \textbf{S}elf-Learning: the facilitation of these exercises enables participants to ask for clarifications when needed, enabling each participant to start from their own knowledge baseline. 

\section{Next steps and Future Work}
The exercise is ready for testing with real-world \acp{CSIRT} and executives. This stage will be used to identify UI bugs, and refine the content of the event and asset database. Once this is complete the exercise will be made freely available under a Creative Commons License for organisations to use. We will continue to work with our industry partners using the exercise to gather data around how organisations prioritise and respond to different incidents and triggers. 

Future work will focus on the creation of new asset and event databases for different sectors, making it possible to explore how different sectors and different technology stacks handle incident scenarios. Decisions and Disruptions has always helped organisations better understand how they make risk decisions: but now they have to \emph{decide harder}.

\section*{Acknowledgments}
We would like to thank Cyber Griffin, the City of London Police and the City of London Corporation for their funding and support.
\bibliographystyle{plain}
\bibliography{references}

\balance

\newpage
\appendix\onecolumn
\FloatBarrier
\section{Example Physical Cards}
\begin{figure}[H]
  \begin{minipage}[t]{.5\linewidth}
      \begin{center}
        \includegraphics[width={\linewidth}]{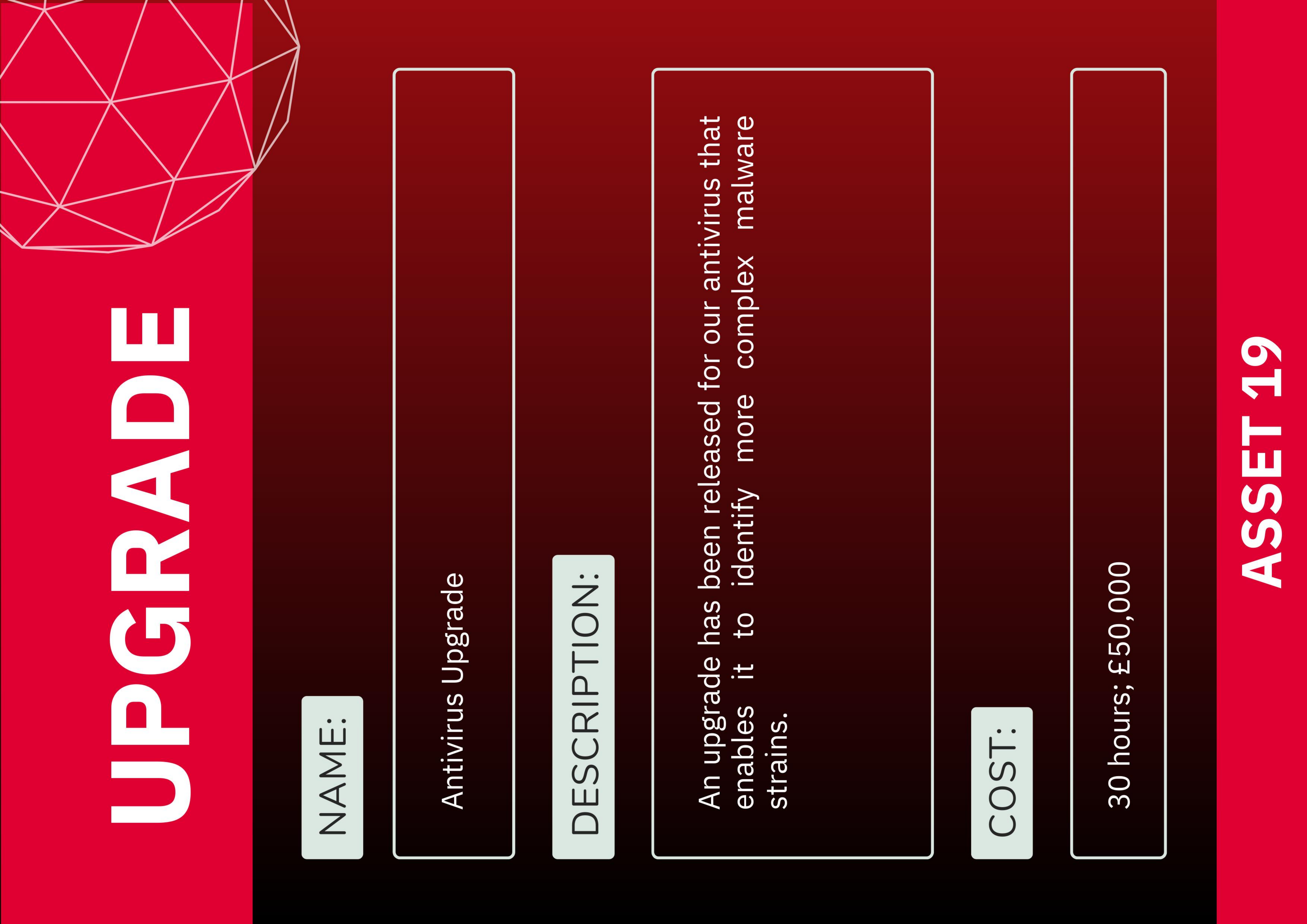}
        \includegraphics[width={\linewidth}]{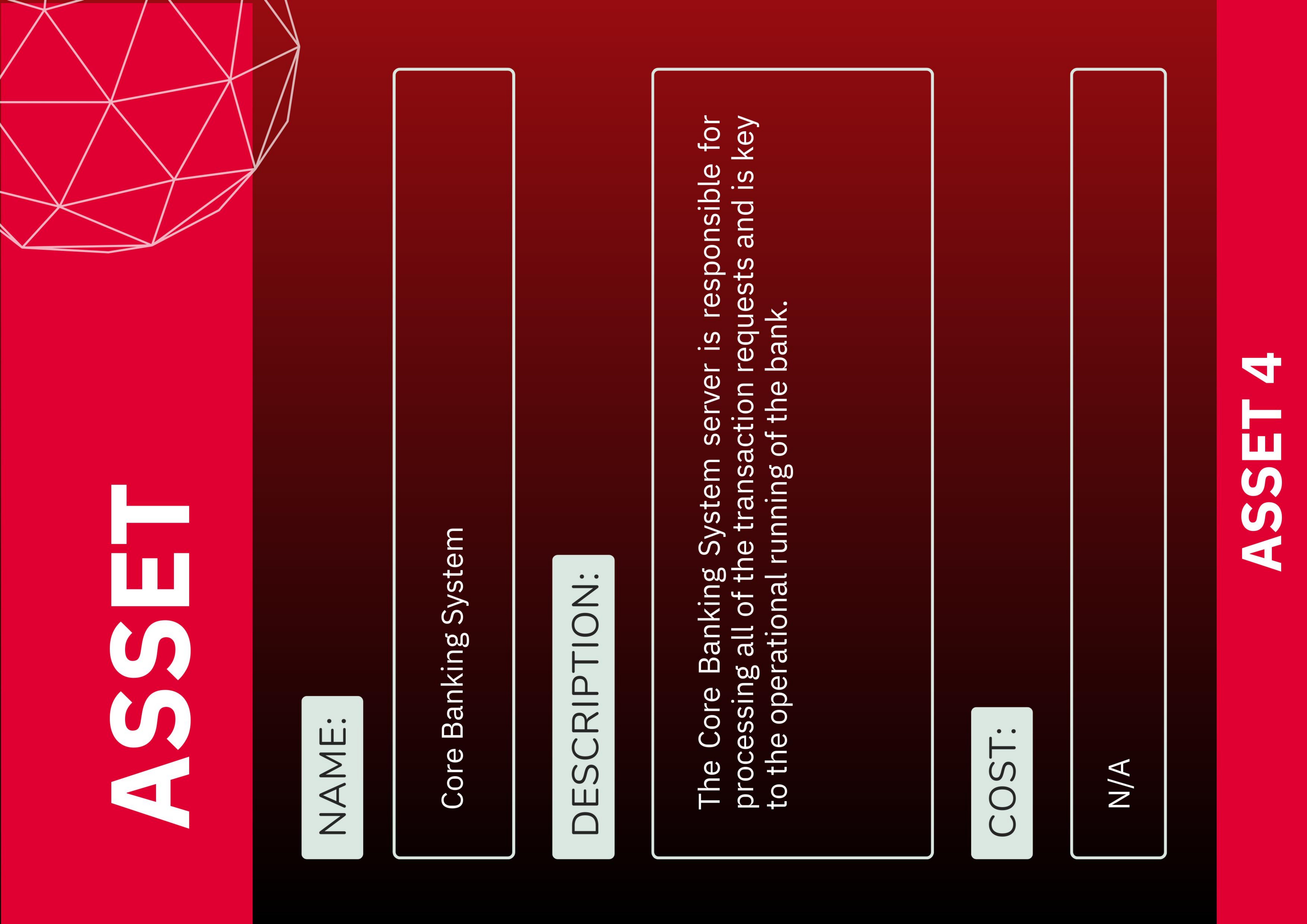}
        \includegraphics[width={\linewidth}]{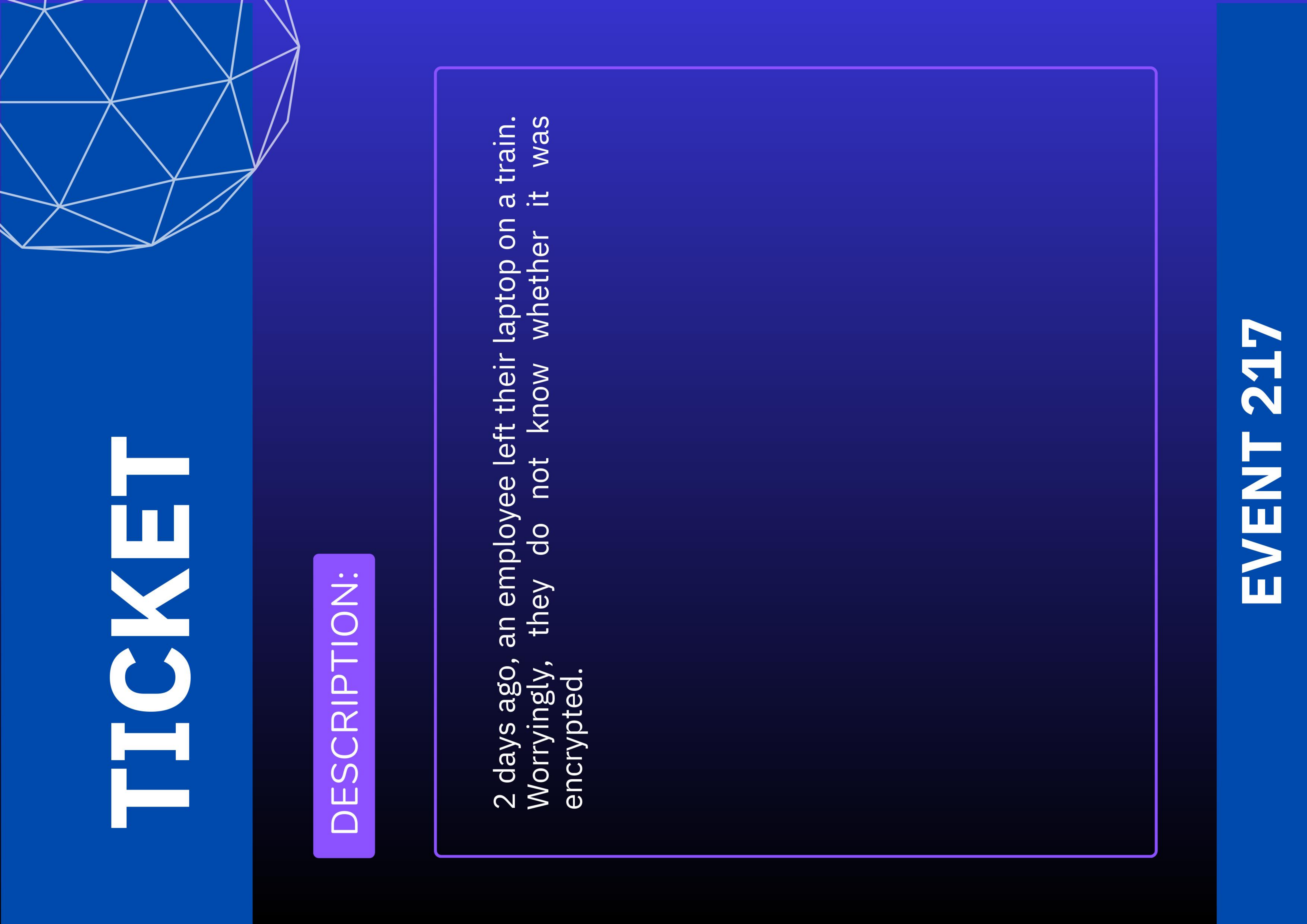}
      \end{center}
    \end{minipage}
    \hfill{}
    \begin{minipage}[t]{.5\linewidth}
        \begin{center}             
          \includegraphics[width={\linewidth}]{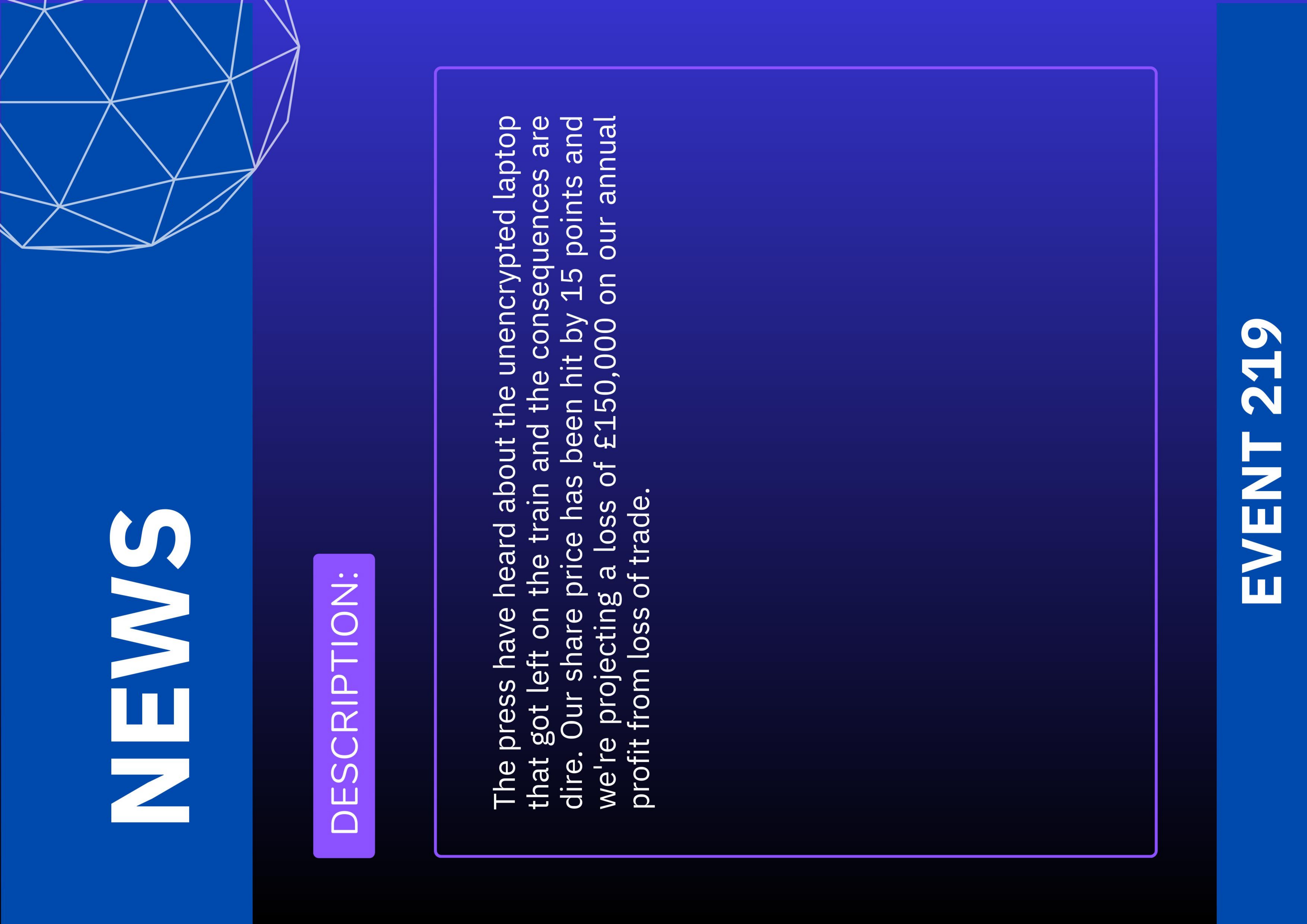}
          \includegraphics[width={\linewidth}]{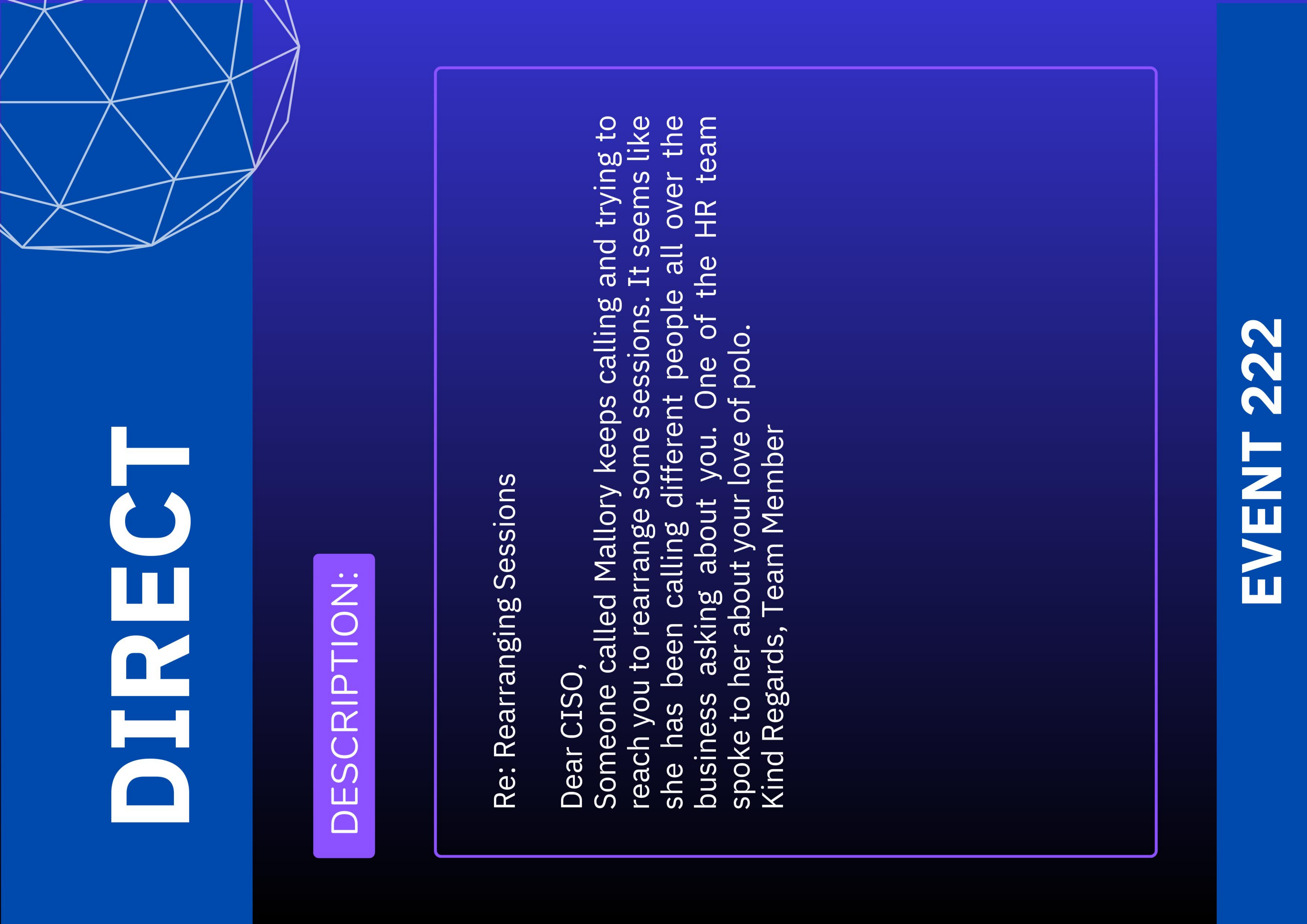}
          \includegraphics[width={\linewidth}]{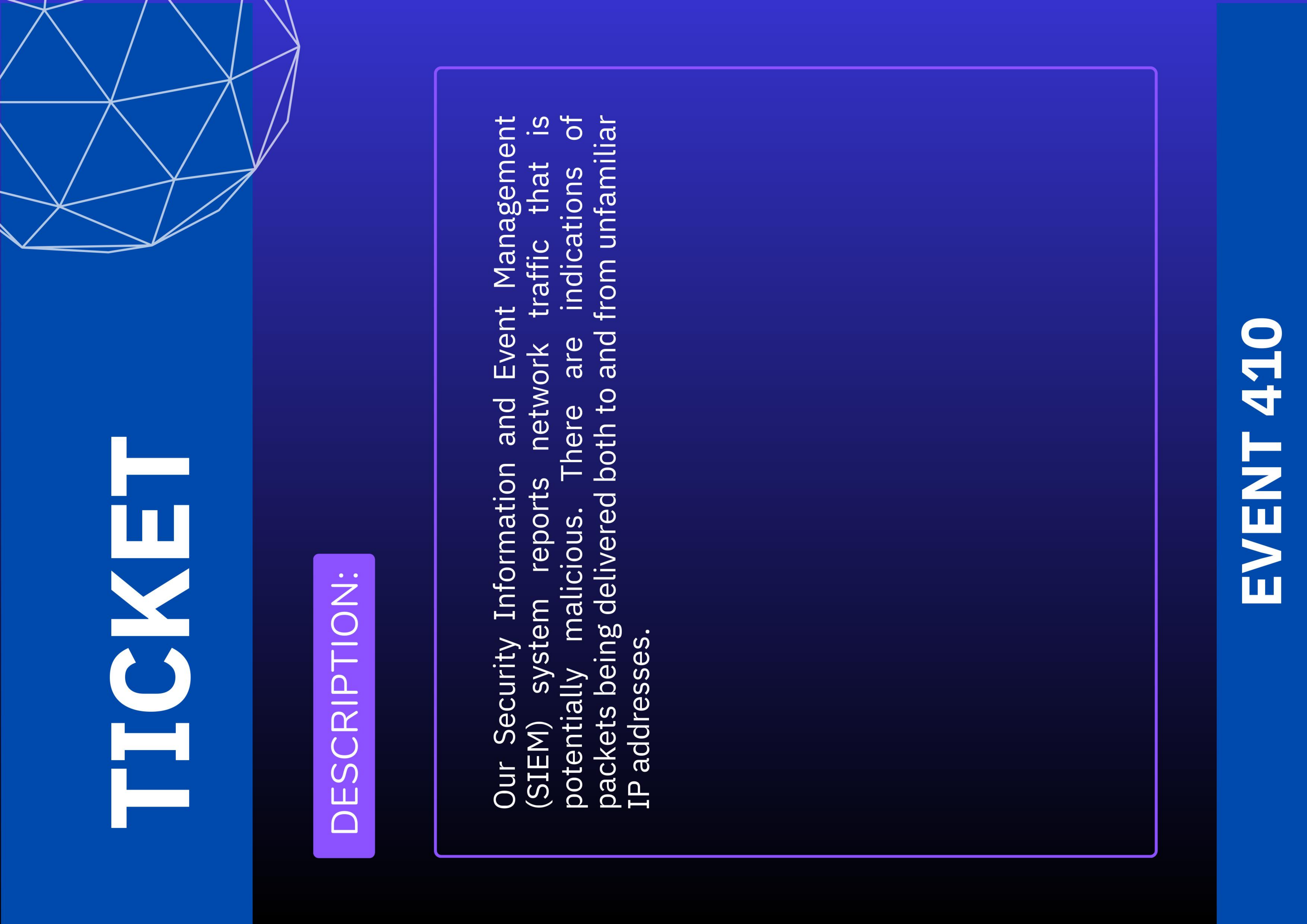}
        \end{center}
    \end{minipage}
  \caption{Examples of physical cards used to play the game when the UI is unavailable.}
  \label{fig:physical-cards}
\end{figure}

\FloatBarrier
\section{Exercise UI}
\label{appx:exerciseUI}
\begin{figure}[H]
    \centering
    \includegraphics[width=.85\linewidth]{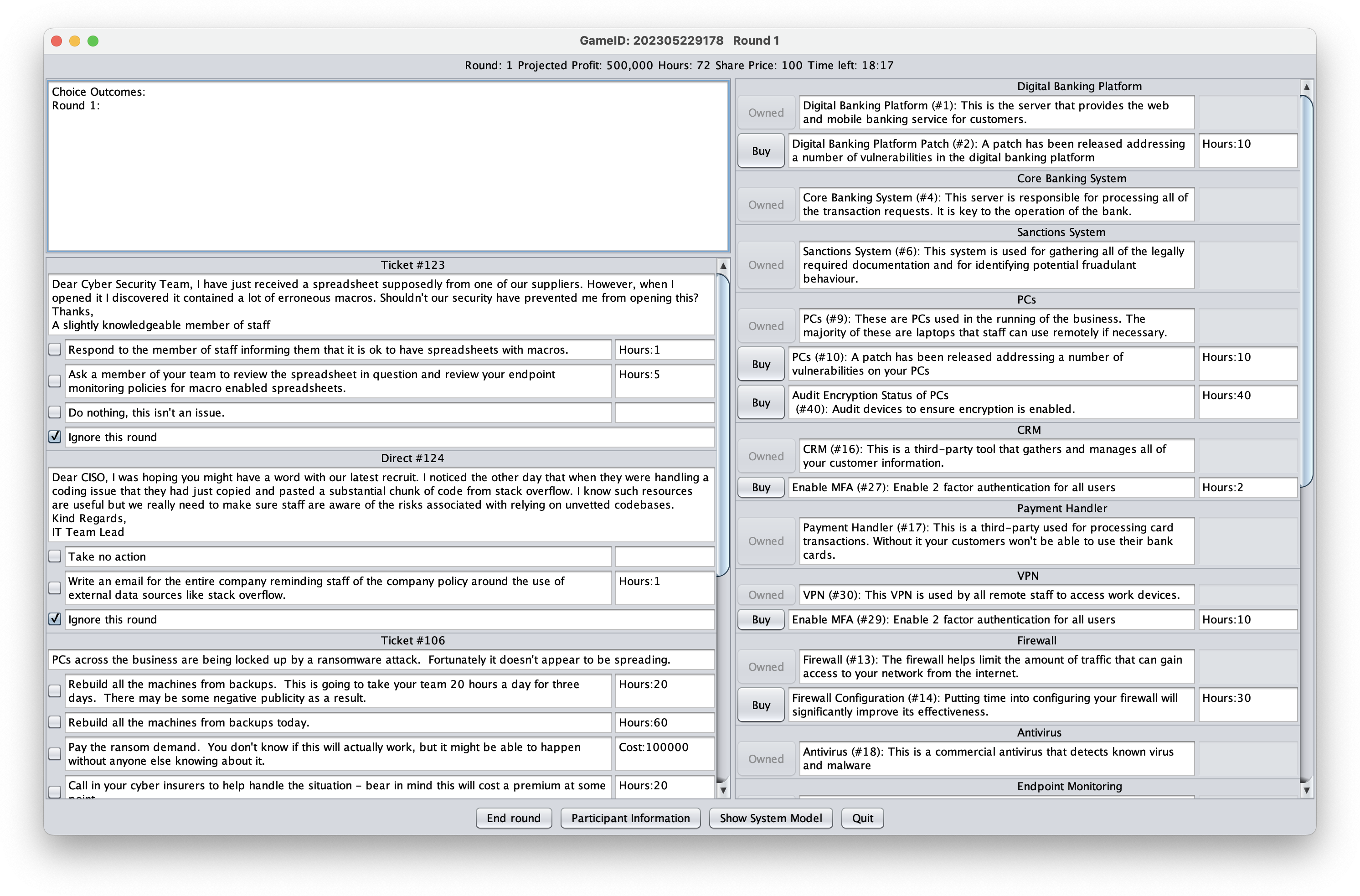}
    \caption{Main exercise interface.}
\end{figure}
\begin{figure}[H]
    \includegraphics[width={\dimexpr .5\linewidth-2\tabcolsep}]{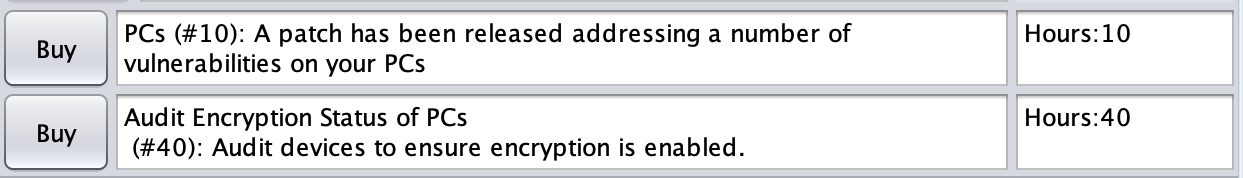}
    \includegraphics[width={\dimexpr .5\linewidth-2\tabcolsep}]{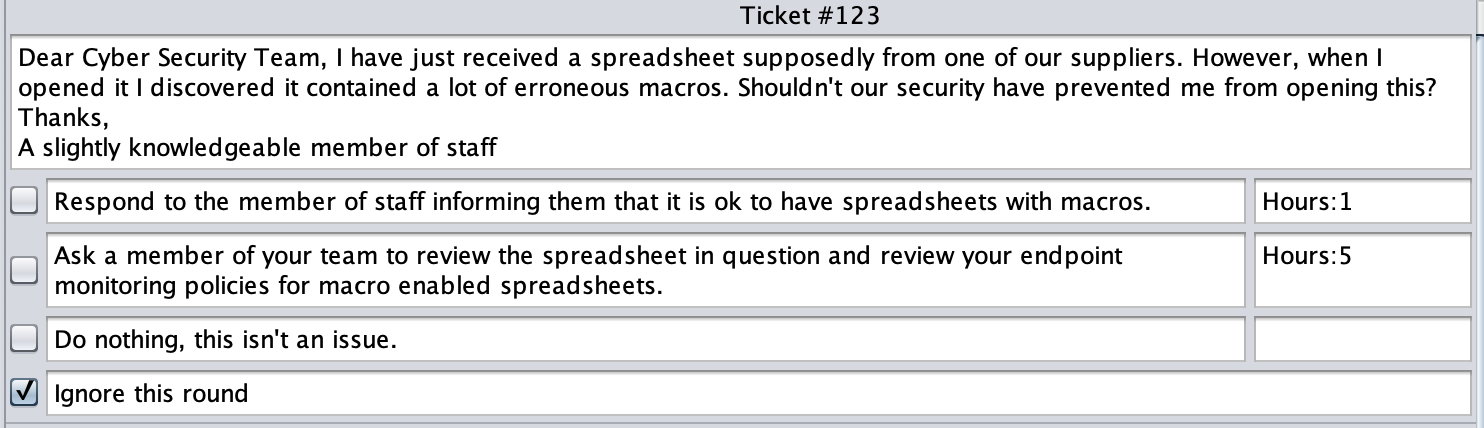}
    \caption{Examples of an event and an \ac{EA}.}
\end{figure}

\end{document}